\documentclass[useAMS,usenatbib]{mn2e}
\usepackage{graphicx}

\title[Star Formation in Mon OB1]
{Star Formation in Massive Protoclusters in the Monoceros OB1 Dark Cloud}

\author[G. Wolf-Chase et al.]
{G. ~Wolf-Chase,$^{1,2}$ 
G. ~Moriarty-Schieven,$^3$
M. ~Fich,$^4$
and
M. ~Barsony,$^5$ \\
$^1$Dept. of Astronomy \& Astrophysics, University of Chicago,
5640 S. Ellis Ave., Chicago, IL 60637\\
$^2$Adler Planetarium \& Astronomy Museum,
1300 S. Lake Shore Drive, Chicago, IL 60605\\
$^3$National Research Council of Canada, 660 N. A'ohoku
Place, Hilo, HI 96720\\
$^4$University of Waterloo, Department of Physics,
Waterloo, ON, N2L 3G1, CA\\
$^5$Space Science Institute, 3100 Marine Street, Suite A353,
Boulder, CO 80303-1058}

\begin{document}

\maketitle

\label{firstpage}

\begin{abstract}

We present far-infrared, submillimetre, and millimetre observations of 
bright {\it IRAS} sources and outflows
that are associated with massive CS clumps
in the Monoceros OB1 Dark Cloud. 
Individual star-forming cores are identified within each clump.
We show that combining submillimetre maps, obtained with SCUBA on the JCMT,
with HIRES-processed and modelled {\it IRAS} data is a powerful technique that
can be used to place better limits on individual source contributions to the 
far-infrared flux in clustered regions.
Three previously categorized ``Class I objects''
are shown to consist of multiple
sources in different evolutionary stages. 
In each case,
the {\it IRAS} point source dominates the flux at 12 \& 25 $\mu$m.
In two cases, the {\it IRAS} point source is not evident at submillimetre
wavelengths.
The submillimetre sources
contribute significantly to the 60 \& 100 $\mu$m fluxes, dominating
the flux in the 100 $\mu$m waveband.
Using fluxes derived from our technique,
we present the spectral energy distribution and physical parameters for an
intermediate-mass Class 0 object in one of the regions.
Our new CO J=2$\rightarrow$1 outflow maps of the three regions studied
indicate complex morphology suggestive of multiple driving sources.
% Where possible, we compare our data with published shorter
% wavelength observations that identify near-infrared sources and Herbig--Haro
% flows in these regions.
We discuss the possible implications of our results for published correlations
between outflow momentum deposition rates and ``source'' luminosities,
and for using these derived properties to estimate
the ratio of mass ejection rates to mass accretion rates onto protostars.

\end{abstract}


\section{Introduction}

% The prevailing models for the formation of isolated, low-mass stars
% all utilize some form of magnetocentrifugal acceleration from circumstellar
% disks that carries off infalling material
% with excess angular momentum (Pudritz \& Norman 1983, 1986;
% Lovelace, Wang, \& Sulkanen 1987; K\"onigl 1989;
% Pelletier \& Pudritz 1992; Wardle \& K\"onigl 1993; Safier 1993; Paatz \&
% Camenzind 1996; Shu et al. 2000).
One of the outstanding questions in star formation is whether high-mass stars
form in a similar manner to low-mass stars, but the answer to this 
question is poorly constrained observationally.
Although
the tendency for stars to form in clusters in regions of intermediate- and
high-mass star formation has been firmly established
(Testi, Palla, \& Natta 1999),
the main database for infrared data on these stellar nurseries
remains that of {\it IRAS}, which typically probes
large cloud structures identified with
the formation of clusters or groups of stars (clumps), rather than the 
smaller condensations
($\sim$0.1 pc $-$ cores) out of which single stars or multiple stellar systems
form (McKee 1999; Myers 1999).

For many years now, the {\it IRAS} database has been used extensively 
to help categorize the spectral energy distributions (SEDs) of young stellar
objects (YSOs) according to the proposed evolutionary scheme developed
by Adams, Lada, \& Shu (1987). Although this scheme was originally proposed
to describe an evolutionary sequence for isolated low-mass star formation,
it has since
been applied to many {\it IRAS} sources that are often high-luminosity
objects, begging the question of what an SED means in cases where it describes
not a single star-forming core, but rather a group of objects, which may 
or may not be coeval.
Furthermore, since the 
earliest surveys for bipolar outflows in molecular clouds (Bally \& Lada
1983; Lada 1985), the
{\it IRAS} database has been used extensively to help determine the 
luminosities of the driving sources of these outflows and to further establish
correlations between source and outflow properties (e.g., Panagia 1991;
Cabrit \& Bertout 1992).
A recent survey for jets in the Vela molecular clouds produced
evidence that many jets are driven by low-mass objects clustered around
{\it IRAS} sources which are likely to be intermediate-mass YSOs
(Lorenzetti et al. 2002).
This raises the interesting question of how the
observed correlations between outflow momemtum deposition rates and ``source''
luminosities should be interpreted.
% since {\it IRAS}-derived luminosities
% provide only 
% upper limits to the luminosities of the source or sources that are actually
% driving the outflows.

Submillimetre continuum imaging with arrays such as SCUBA (Submillimetre
Common-User Bolometer Array)
on the 15-m James Clerk Maxwell Telescope on Mauna Kea can probe conditions
on the scale of individual star-forming cores, and thus
can be useful in separating sources in confused regions, but by
themselves, these observations yield limited
spectral coverage for the determination of physical properties such as 
temperature, bolometric luminosity, and evolutionary stage. Since protostars
have SEDs that peak in the far-infrared, multiband
infrared data are critical, though difficult to obtain given the
current lack of operational far-infrared observatories.
However, software advances in HIRES (HIgh-RESolution)
processing of {\it IRAS} data (Aumann, Fowler, \& Melnyk 1990)
have allowed unprecedented spatial resolutions to be achieved at 
FIR wavelengths
(Surace et al. 1993; Terebey \& Mazzarella 1994; Cao et al. 1996; Hurt \&
Barsony 1996; Cao et al. 1997; Barsony et al. 1998).
% \citep{sur93,ter94,cao96,hur96,cao97,bar98}.
HIRES-processing and modelling of {\it IRAS} data has even been
suggested as a method for
identifying Class 0 protostars (O'Linger et al. 1999).

% On the one hand, there is some evidence that young high-mass stars have
% associated disks and possibly highly-collimated jets and outflows (e.g.,
% Zhang et al. 2002) similar to low-mass stars. 

In this paper, we apply HIRES processing and modelling techniques
to massive star-forming clumps in
the Monoceros OB1 dark cloud.
Molecular emission from the Mon OB1 dark cloud was first mapped
by Blitz (1978)
at low resolution covering 11 square degrees towards the arc of dark nebulae
seen on the Palomar Observatory Sky Survey print (Lynds 1962).
Since the Mon OB1 dark cloud lies toward the outer Galaxy,
near the galactic anticentre,
confusion due to foreground and background clouds is at
a minimum, affording an excellent opportunity to study the large-scale
distribution of molecular material and associated star formation.
% The Monoceros OB1 dark cloud was first identified by Barnard (1919) in a
% systematic search for dark patches on the night sky, but little work was
% done on the detailed structure and evolution of dark clouds until the
% advent of millimeter-wave astronomy in the 1970s.
% Today, millimeter emission
% from CO has been mapped over the entire Galactic plane, as well as over many
% high-latitude clouds (Dame, Hartmann, \& Thaddeus 2001).
Oliver, Masheder, \& Thaddeus (1996, hereafter OMT96) performed
an unbiased CO survey over 52 square degrees at higher
sensitivity and spectral resolution than the Blitz survey
in order to gain a better understanding of the large-scale
structure and kinematics of the ISM towards the Mon OB1 region and study the
properties of molecular clouds in the outer Galaxy.
They used the {\it IRAS} Point Source Catalogue to identify sites of massive
star formation in their large-scale CO survey by applying
two criteria using color selection of {\it IRAS} point sources
(Richards et al. 1987; Wood \& Churchwell 1989).

The star-forming region that OMT96 identify as Cloud 16,
associated with the young star cluster
NGC 2264, has been the target of many molecular line 
studies, including
an unbiased CO J=1$\rightarrow$0
survey for molecular outflows (Margulis, Lada, \& Snell 1988,
hereafter MLS88)
and search for dense gas via a multitransitional CS
study (Wolf-Chase, Walker, \& Lada 1995, hereafter WWL95;
Wolf-Chase \& Walker 1995, hereafter WW95).
Prior to publication of the {\it IRAS} point source catalog, Margulis, Lada,
\& Young (1989; hereafter MLY89)
used the {\it IRAS} database to identify 30 discrete sources
within this region,
and classify their SEDs in terms of the 
evolutionary scheme proposed by Adams, Lada, \& Shu (1987).
Four of these sources lie at or near the emission peaks of the CS clumps
identified by WWL95:
IRAS 9 ({\it IRAS}
06382$+$0932; also NGC 2264 IRS 1 $-$ Allen
1972),
IRAS 12 ({\it IRAS} 06382$+$0939),
IRAS 25 ({\it IRAS} 06382$+$1017), 
\& IRAS 27 ({\it IRAS} 06381$+$1039).
Three of these sources are also identified as sites of massive star formation by
OMT96 
({\it IRAS} 06382$+$0939,
{\it IRAS} 06382$+$1017, 
\& {\it IRAS} 06381$+$1039).
All three of these
{\it IRAS} sources were categorized as Class I
objects by MLY89.
% There is only one known low-mass object in Mon OB1 that satifies the
% criteria established by Andr\'e, Ward-Thompson, \& Barsony (1993) for
% Class 0 protostar status:
% {\it IRAS} 06384$+$0958,
% the driving source of the highly-collimated outflow NGC 2264 G 
% (Ward-Thompson, Eiroa, \& Casali 1995; Margulis et al. 1990; Lada \& Fich
% 1996).
% This source lies well outside of the massive CS cores,
% in an apparently isolated region of the cloud.
% The isolation of this source stands out in stark contrast to other regions
% of low-mass star formation, such as Serpens and Perseus,
% which are forming clusters
% (Hurt \& Barsony 1996; Barsony et al. 1998; Wolf-Chase, Barsony, \& O'Linger
% 2000).

The primary goal of this paper is to identify individual star-forming cores
located in the massive CS clumps in Mon OB1 and estimate their contributions
to the {\it IRAS} fluxes that characterize the entire clumps, in order to
assess the accuracy of source classes and luminosities
that were previously deduced from the {\it IRAS} data.
% Where possible, we
% identify the driving sources of molecular outflows and Herbig-Haro flows
% in these regions.
We concentrate on the three sites of massive star formation
identified by OMT96, since the fourth {\it IRAS} source that coincides with a 
CS emission peak, NGC 2264 IRS 1,
has already been studied in great detail in both molecular line
and continuum emission
(e.g., Schreyer et al. 1997; Ward-Thompson
et al. 2000; Wang et al. 2002).
In addition to the {\it IRAS} Point Source Catalogue designations,
we use the nomenclature of MLS88 \& MLY89
to refer to outflows and {\it IRAS}
sources in this cloud.
We adopt a distance to Mon OB1 of 800 pc (Walker 1956); the same distance
adopted by MLS88 and MLY89. This is a compromise between
more recent estimates of 700 $\pm$
                                 40 pc and 790 pc given by Feldbrugge
\&  van Genderen (1991) and Sagar \&  Joshi (1983) 
                      based on VBLUW and UBV photometry of stars in the
young cluster NGC 2264, and 950 pc, adopted by OMT96, based on 
photometric analysis by P\'erez, Th\'e, \& Westerlund (1987).

\section
{Observations and Data Reduction}

\subsection
{Submillimetre Continuum Mapping: JCMT Observations}

The observations presented here were obtained 
with the Submillimetre Common-User
Bolometer Array (SCUBA) (Holland et al. 1999)
at the James Clerk Maxwell
Telescope (JCMT)\footnote{The JCMT 
is operated by the Joint Astronomy Centre on
behalf of the Particle Physics and Astronomy Research Council
of the United Kingdom, the Netherlands
Organization for Scientific Research, and the National Research Council of
Canada.} located near the summit of Mauna Kea, Hawaii. 
The instrument contains two arrays of
bolometric detectors -- the Short Wave (SW) array has 91 pixels optimised
for observations at 450 $\mu$m, and the Long Wave (LW) array has 37
pixels at 850 $\mu$m.  Both arrays can be used simultaneously and have
approximately a 2.3 arcmin diameter field-of-view on the sky. The
instrument achieves sky-background limited performance by cooling the 
detectors to $\sim$75 mK.

The data were obtained 9 August, 18 and 23 November 1999, and 21
October 2000, using the standard SCUBA ``scan-map'' observing mode
(Jenness et al. 2000). 
Fully sampled maps at both 450 $\mu$m and 850 $\mu$m were generated with
3 arcsec sampling.  The pointing was checked using a nearby blazar every
$\sim$1 hour, and was found to vary by $<$2 arcsec.  The data were
corrected for atmospheric extinction by the method recommended by
Archibald et al. (2002), i.e. fitting a polynomial to the 225 GHz zenith
optical depth derived by a tipping radiometer, then extrapolating to
850 $\mu$m and 450 $\mu$m.  The data were reduced and maps reconstructed
using the ORAC data reduction package (Jenness \& Economou 2001).

Flux calibration was applied to the data using images of a planet or
of the secondary calibrator source CRL618, obtained using the same
observing technique as the data.  Flux densities of the sources were
derived using aperture photometry with a 45 arcsec diameter aperture,
and then correcting for flux in the error beam using the same flux
calibration images.

\subsection
{NRAO 12-m Continuum Photometry}

We obtained 1.3 mm continuum data 
during December 1997, using the National Radio Astronomy Observatory
(NRAO) 12-m telescope located on Kitt Peak, near Tucson, 
Arizona\footnote{The National Radio Astronomy Observatory
is a facility of the National Science Foundation, operated under cooperative
agreement by Associated Universities, Inc. The 12-m telescope 
(currently known as the Arizona
Radio Observatory) is now operated through the University of Arizona's
Steward Observatory, with funds provided by the University of Arizona, the
Research Corporation, and the National Science Foundation.}
toward the position of three
compact submillimetre sources that we identified in earlier,
unpublished, 850 $\mu$m SCUBA jiggle-maps, which we acquired prior to the 
scan-maps reported in this paper.
We measured the 1.3 mm continuum flux toward these sources within a
27 arcsec beam using a dual-channel, double-sideband,
SIS heterodyne receiver system.  The receiver had a bandwidth of 600 MHz
and was operated at a sky frequency of 231.6 GHz.  The subreflector was nutated
at a frequency of 4 Hz, using a 2 arcmin beam throw.  Data were calibrated
by chopping between sky and an ambient temperature load.  Due to the inopportune
positions of the planets, absolute calibration was achieved by observing
NGC 2071 IR and NGC 2264 IRS,
and using published 1.3 mm fluxes to scale our data 
(Walker, Adams, \& Lada 1990; Sandell 1994).
Consequently,
we estimate 1.3 mm flux uncertainty to be at the 30\%--40\% level.

\subsection
{Far-Infrared Mapping: HIRES Processing of the {\it IRAS} Data}

A detailed background of {\it IRAS} data and HIRES processing 
techniques may be found in Hurt \& Barsony (1996), Barsony et al. (1998),
and O'Linger et al. (1999), including an in-depth discussion of the
significant merits of taking HIRES-processing beyond the standard 
20 iterations, which is the default when HIRES-processing is requested
remotely via electronic mail to IPAC.\footnote{IPAC is funded by NASA as
part of the {\it IRAS} extended mission under contract to JPL.}
For the data presented here, we halted HIRES processing at 40 iterations in the
12 $\mu$m and 25 $\mu$m bands, and at 200 iterations in the 60 $\mu$m and 100
$\mu$m bands.  These images do not represent ``converged'' data in a formal 
sense, as the concept of convergence in iterative solutions of nonlinear
problems seldom has any absolute meaning (O'Linger et al. 1999).
Our iteration limits 
in the various wavebands were chosen by looking at the change in the correction 
factor variance from one iteration to the next, and setting an acceptable 
threshold value (Aumann et al. 1990; Hurt \& Barsony 1996; 
O'Linger 1997; Barsony et al. 1998).

The irregular sampling of the sky brightness distribution by subsequent passes
of the {\it IRAS} detectors results in variable spatial resolution across a 
HIRES-processed image.
Among the diagnostic maps produced during the HIRES-processing, are 
``beam-sampled'' maps, which may be used to compute effective resolutions
achieved at various positions in the HIRES-processed image by inputting 
a series of user-defined, point-source
``spikes'' across the image. These spikes are then
HIRES-processed along with the actual {\it IRAS} data,
allowing effective resolutions
to be calculated at the positions of the spikes.
Ideally, the beam-sampled maps can be used to produce models of the actual 
HIRES-processed {\it IRAS} emission. 
For example, supplying user-defined spikes to
coincide in magnitude and position with actual sources chosen from the
{\it IRAS}
Point Source Catalogue results in a point-source model, which can be compared
to the actual HIRES-processed data (Hurt \& Barsony 1996).
One may subsequently add further spikes to more
closely simulate the actual emission (O'Linger 1997; Barsony et al. 1998).
Such spikes may represent
extended emission, or, in some cases, additional embedded sources.
Point-source
modelling offers significant advantages over simple software aperture
photometry of the HIRES-processed {\it IRAS} survey images. Whereas simple
software aperture photometry suffers from uncertainties as to the proper
aperture size to use, and the systematic effects that can be introduced
by the processing algorithm itself, the input fluxes for the best-match image
to the {\it IRAS} survey data in point-source models are known.

In some cases, higher-resolution observations may be available at a different
wavelength, which can be used to estimate placement of input spikes.
Our high-resolution (14 arcsec at 850 $\mu$m;
7 arcsec at 450 $\mu$m)
SCUBA observations have given us prior knowledge of sources that are 
embedded in dense cores in the Mon OB1 dark cloud. We have used this prior
knowledge to model the HIRES-processed {\it IRAS} emission from these sources.
The default pixel size of 15 arcsec was used.
Absolute calibration uncertainty for HIRES-derived fluxes is estimated to be 
20\% (Levine \& Surace 1993).

\subsection
{NRAO 12-m OTF Mapping}

We acquired CO J$=$2$\rightarrow$1 maps of outflows
NGC 2264 O \& NGC 2264 H 
(associated with MLY89 sources IRAS 25, \& IRAS 27)
% at twice the resolution of 
% previously-available CO J$=$1$\rightarrow$0 maps
% (MLS88),
using the spectral-line On-The-Fly (OTF) mapping mode
of the NRAO's 12-m telescope in March 1996, and of NGC 2264 D (associated
with MLY89 source IRAS 12) in May 1999.
The OTF technique allows the acquisition of large-area,
high-sensitivity, spectral line maps with unprecedented speed and pointing 
accuracy. 

% Although OTF mapping is not a new concept, 
% given the rigor of the position encoding that allows precise and accurate
% gridding of the data, the fast data recording rates that allow rapid scanning 
% without beam smearing, and the analysis tools that are available,
% the 12m implementation is the most ambitious effort at OTF imaging yet.

A dual-channel, single-sideband SIS receiver was used for all observations.
For the March 1996 observations, the backend consisted of a 1536-channel hybrid
spectrometer. The 1536 channels were divided among the two receiver 
polarization channels. A bandwidth of 600 MHz, yielding a spectral resolution
of 781 kHz (1 km s$^{-1}$), was used for NGC 2264 O, and a bandwidth of 300
MHz, yielding a spectral resolution of 391 kHz (0.5 km s$^{-1}$), was used for
NGC 2264 H.
For the May 1999 observations of NGC 2264 D, the backend consisted of 
500 kHz and 1 MHz resolution filterbanks, yielding velocity resolutions of
0.65 km s$^{-1}$ and 1.3 km s$^{-1}$, respectively. The filterbanks 
were used in parallel mode, each of the two receiver polarization channels
using 256 filterbank channels.
The polarization channels were subsequently averaged
together to improve signal-to-noise.
Only the 500 kHz resolution data were used to produce the final map.
We performed 5 map coverages for each outflow, attaining an RMS
$\sim$0.25--0.35 K for the added maps.
We used the NRAO standard source Orion A
($\alpha_{1950}=$05$^h$ 32$^m$ 47.0$^s$, 
$\delta_{1950}=$-05$^{\circ}$ 24$^{\prime}$ 21$^{\prime\prime}$)
to check absolute line temperatures.

The OTF data were reduced with the Astronomical Image Processing
Software (AIPS), Version 15JUL95.
AIPS tasks specific to OTF data 
are `OTFUV', which converts a single 12-m OTF map (in UniPOPS SDD
format) to UV (single-dish) format,
and `SDGRD', which selects random position single-dish data in AIPS UV format
in a specified field of view about a specified position and projects the
coordinates onto the specified coordinate system.
The data are then convolved onto a grid.
OTF data maps were first combined, then gridded into a
data cube and baseline-subtracted.
Channel maps as well as
individual spectra were inspected
to ensure good baseline removal and to check for scanning artifacts.

\section
{Results}

\subsection{Background}

\subsubsection{{\it IRAS} 06382+0939 (IRAS 12)}

IRAS 12 was classified as a Class I object by MLY89.
Identified as IRAS 06382$+$0939 upon publication of the Point Source
Catalogue, it
is embedded in the northern part of
a massive
(1900--2500 M$_{\odot}$) clump traced by CS J=2$\rightarrow$1 emission 
(WWL95).
J, H, K, \& L$^{\prime}$ maps of the
{\it IRAS} source revealed two point sources (RNO--East \& RNO--West)
separated by 2.8 arcsec
along a position angle of 265 degrees
at the PSC position (Castelaz \& Grasdalen 1988).
The effective temperatures of RNO--East \& RNO--West were found
to be 3000 K and 10,000 K, respectively, with a combined luminosity of
about 550 L$_{\odot}$.
The age of RNO--East was found to be less than 10$^5$ years old, and
RNO--West was found to be a young high-mass star (Castelaz \& Grasdalen 1988).
% A K$^{\prime}$ survey by
% Hodapp (1994) 
% revealed extended nebulosity in the vicinity of these near-infrared sources.
This region was also previously observed in the far-infrared at 70 \&
130 $\mu$m (Sargent et al. 1984) and from 40 to 160 $\mu$m (Cohen,
Harvey, \& Schwartz 1985). The latter reported a double peak, based on
the higher-resolution (40--45 arcsec)
capabilities of the Kuiper Airborne Observatory (KAO):
their primary peak corresponds to the position of the {\it IRAS} PSC position,
while their secondary peak lies more than 2 arcmin to the southeast,
approximately at the peak of the CS J=2$\rightarrow$1 and CS J=5$\rightarrow$4
emission reported by WWL95.

MLS88 associated IRAS 12 with a large
($\approx$12 arcmin in extent), massive (16--30 M$_{\circ}$),
bipolar, CO outflow (NGC 2264 D), oriented with its major axis
along a NE--SW direction, that has a dynamical
time-scale of $\sim6.9\times10^4$ yr.
High-velocity wings associated with the outflow were also observed in 
CS J=2$\rightarrow$1 spectra (WW95).
The outflow centroid is displaced 
$\sim$2 arcmin from the {\it IRAS} PSC position.
A weak ($\approx$ 0.6 mJy) VLA 6 cm source;
% approximately 30 arcsec north-east of {\it IRAS} 06382$+$0939, a
a strong H$_2$O maser;
% at $\alpha(1950)=06^h38^m22.8^s$, 
% $\delta(1950)=09^{\circ}36^{\prime}35^{\prime\prime}$
and four near-infrared sources all lie within the boundaries of the outflow,
distributed in a band approximately perpendicular to the outflow axis.
One of the near-infrared sources (IRS A) was identified as the probable driving
% IRS C: the near-infrared counterpart to {\it IRAS} 06382$+$0939,
% IRS B: associated with the star W187 ($=$SAO 114267, a B7 V star), 
% IRS A: associated with a star-like object W166, probably a
% Herbig Be/Ae star, 
% IRS F: located near the H$_2$O maser and identified as the possible H$_2$O 
% maser driving source.
source of the
outflow, based on its location closest to the outflow centroid of all the
near-infrared sources (Mendoza et al. 1990).

\subsubsection{{\it IRAS} 06382+1017 (IRAS 25)}

IRAS 25 was classified as a Class I object by MLY89.
Identified as IRAS 06382$+$1017 upon publication of the Point Source
Catalogue, it lies
at the southern end of an extended, massive (500--700
M$_{\odot}$) clump traced by CS J=2$\rightarrow$1 emission (WWL95). 
It has been associated with a
compact CO outflow,
NGC 2264 O (WWL95; WW95);
a giant Herbig--Haro flow,
HH124, and its bow-shock pairs, HH124--E \& HH124--W; an infrared
reflection nebula; and a near-infrared
source
(Walsh, Ogura, \& Reipurth 1992; Moneti \& Reipurth 1995;
Ogura 1995; Pich\'e, Howard, \& Pipher 1995).
The near-infrared source is a barely resolved binary,
with the secondary component located
$\sim$1.75 arcsec ($\sim$1400 AU)
from the primary at a P.A.$\approx$155 degrees.
The Herbig--Haro objects lie along a P.A.$\approx$105 degrees, while the 
infrared reflection nebula lies along a P.A.$\approx$45 degrees. Pich\'e
et al. (1995) suggested that the infrared reflection nebulosity might be due to
the lobes of a second outflow cavity extending toward the
north-east, or, alternatively, reflection off a circumstellar torus
whose polar axis is roughly parallel to the axis of the Herbig--Haro flow.
Rodr\'iguez \& Reipurth  (1998) reported two VLA sources within the error
ellipse for {\it IRAS} 06382$+$1017 (VLA1: $\alpha(1950)=06^h38^m17.01^s$,
$\delta(1950)=+10^{\circ}17^{\prime}56.2^{\prime\prime}$;
VLA2: $\alpha(1950)=06^h38^m17.91^s$,
$\delta(1950)=+10^{\circ}17^{\prime}58.3^{\prime\prime}$).
VLA 1 is approximately coincident with the position of the near-infrared
source and reflection nebula.

\subsubsection{{\it IRAS} 06381+1039 (IRAS 27)}

IRAS 27 was classified as a Class I object by MLY89.
Identified as IRAS 06381$+$1039 upon publication of the Point Source
Catalogue, it
lies near the northern end
of a massive (500--700
M$_{\odot}$) clump traced by CS J=2$\rightarrow$1 emission (WWL95).
MLS88 associated this source with an
outflow that they identified as having 
a shorter axis along the direction of the outflow than 
perpendicular to the direction of the outflow (NGC 2264 H).
The {\it IRAS} source
is offset about 45 arcsec west of the apparent outflow centre.
A striking asymmetry is seen in the velocity extents of the blue- and 
redshifted gas; the redshifted gas is evident from 10--30 km s$^{-1}$, while
the blueshifted gas was seen only from 0--3 km s$^{-1}$.
The mass of this outflow was computed to be
M$_{flow}=1.6--2.3$ M$_{\circ}$; and its
dynamical time-scale,
$\tau_d=1.2\times10^4$ yr,
makes it the second youngest of the nine outflows mapped by MLS88.
High-velocity wings associated with the outflow were also observed in 
CS J=2$\rightarrow$1 spectra (WW95).

\subsection{SCUBA images}

In this section, we present our 450 \& 850 $\mu$m SCUBA maps of the
{\it IRAS} sources. Table 1 lists 450 \& 850 $\mu$m fluxes for all
the submillimetre sources that were identified in the three regions, the
1.3 mm fluxes that were obtained for a few sources
at the former NRAO 12-m telescope on
Kitt Peak, and the errors associated with the fluxes.

\begin{table*}
\centering
\begin{minipage}{126mm}
 \caption{Mon OB1 SCUBA \& Millimeter Source Fluxes}
 \begin{tabular}{@{}lccccc@{}}
  \hline
  Source  &  $\alpha(2000)$  &  $\delta(2000)$  & 
  450 $\mu$m
  &  850 $\mu$m
  &  1300 $\mu$m  \\
        &         &       &  (Jy) &  (Jy)  &  (Jy)  \\
  \hline
  RNO  & 
  06$^h$41$^m$02.8$^s$   & 09$^{\circ}$36$^{\prime}$10$^{\prime\prime}$
  &  $<$29.6
  &  $<$1.86  &  \dots  \\
  12 S1 & 
  06$^h$41$^m$05.8$^s$   & 09$^{\circ}$34$^{\prime}$09$^{\prime\prime}$
  &  47.9$\pm$0.51  &  5.38$\pm$0.019 &  1.4$\pm$0.49  \\
  12 S2 & 
  06$^h$41$^m$06.2$^s$   & 09$^{\circ}$35$^{\prime}$57$^{\prime\prime}$
  &  57.0$\pm$0.51 &  5.35$\pm$0.019 &  \dots  \\
  12 S3 & 
  06$^h$41$^m$04.1$^s$   & 09$^{\circ}$35$^{\prime}$01$^{\prime\prime}$
  &  47.9$\pm$0.51 &  4.94$\pm$0.019 &  
  \dots \\
  12 S4 & 
  06$^h$41$^m$00.9$^s$   & 09$^{\circ}$35$^{\prime}$28$^{\prime\prime}$
  &  42.1$\pm$0.51 &  4.15$\pm$0.019 &  \dots  \\
  12 S5 & 
  06$^h$40$^m$49.3$^s$   & 09$^{\circ}$34$^{\prime}$36$^{\prime\prime}$
  &  16.20$\pm$0.51 &  2.171$\pm$0.019 &  \dots  \\
  12 S6 & 
  06$^h$40$^m$57.8$^s$   & 09$^{\circ}$36$^{\prime}$25$^{\prime\prime}$
  &  32.34$\pm$0.51 &  2.986$\pm$0.019  &  \dots  \\
  12 S7 &
  06$^h$41$^m$11.6$^s$   & 09$^{\circ}$35$^{\prime}$34$^{\prime\prime}$
  &  26.40$\pm$0.51  &  2.652$\pm$0.019 &  \dots  \\
  12 S8 & 
  06$^h$41$^m$09.2$^s$   & 09$^{\circ}$33$^{\prime}$01$^{\prime\prime}$
  &  9.04$\pm$0.51  &  1.447$\pm$0.019  &  \dots  \\
  25 NIR & 
  06$^h$41$^m$02.6$^s$  & 10$^{\circ}$15$^{\prime}$02$^{\prime\prime}$
  & $<$2.2
  & $<$0.056 &  \dots  \\
  25 S1 & 
  06$^h$41$^m$03.5$^s$   & 10$^{\circ}$15$^{\prime}$10$^{\prime\prime}$
  & 7.1$\pm$2.3 & 2.92$\pm$0.024  &  \dots  \\
  25 S2 &
  06$^h$41$^m$04.9$^s$   & 10$^{\circ}$14$^{\prime}$55$^{\prime\prime}$
  & 14.8$\pm$2.3 & 3.33$\pm$0.024  &  \dots  \\
  27 S1 & 
  06$^h$40$^m$58.5$^s$   & 10$^{\circ}$36$^{\prime}$54$^{\prime\prime}$
  &  31.8$\pm$1.6  &  3.20$\pm$0.018   &  0.8$\pm$0.28  \\
  27 S2 & 
  06$^h$40$^m$59.1$^s$   & 10$^{\circ}$36$^{\prime}$09$^{\prime\prime}$
  &  12.1$\pm$1.6  &  1.87$\pm$0.018  &  0.7$\pm$0.25  \\
  27 S3 & 
  06$^h$41$^m$02.0$^s$   & 10$^{\circ}$35$^{\prime}$30$^{\prime\prime}$
  &  12.9$\pm$1.6 & 1.72$\pm$0.018 & \dots  \\
  27 S4 & 
  06$^h$40$^m$54.4$^s$   & 10$^{\circ}$33$^{\prime}$27$^{\prime\prime}$
  & 7.4$\pm$1.6 &  0.65$\pm$0.018 &  \dots  \\
  \hline
 \end{tabular}

 \medskip

 The 450 \& 850 $\mu$m fluxes are integrated over a 45 arcsec
 aperture. The 1300 $\mu$m fluxes represent fluxes within a 27 arcsec
 beam. The 450 \& 850 $\mu$m fluxes for RNO and 25 NIR are given as upper
 limits, since these sources have no corresponding submillimetre peaks.

\end{minipage}
\end{table*}

\subsubsection{{\it IRAS} 06382+0939 (IRAS 12)}

Figure 1a shows the 850 $\mu$m emission associated
with IRAS 12. Near-infrared sources (stars), the VLA source (triangle),
the position of a H$_2$0 maser (box)
identified by Mendoza et al. (1990), and
the {\it IRAS} point source error ellipse,
are indicated. We detected eight compact submillimetre sources (crosses) in
the mapped region, which we designate as ``12 S\#'' in this paper.
Recently, lower-resolution 870 $\mu$m observations of this region resulted in
the detection of seven compact sources (Williams \& Garland 2002).
The map shows extended as well as compact
emission, but there is clearly little emission at the position of the {\it IRAS}
source, corresponding to the RNO binary. The source 12 S1 is coincident with
the other far-infrared peak reported by Cohen et al.
(1985).
It is also approximately coincident with the peaks of the CS J=2$\rightarrow$1
and CS J=5$\rightarrow$4 emission reported by WWL95.
The overall morphology of the extended 850 $\mu$m emission 
is very similar to the morphology of the
CS J=5$\rightarrow$4 emission mapped by WWL95 (see fig. 5b WWL95).
Notably, WWL95 detected no CS J=5$\rightarrow$4 emission at the
position of the {\it IRAS} point source.
Figure 1b shows the 450 $\mu$m emission associated with
IRAS 12. Although significantly
noisier than the 850 $\mu$m data, all of the submillimetre sources
present in the 850 $\mu$m map are identifiable at 450 $\mu$m as well.
The close association of the submillimetre and CS J=5$\rightarrow$4 emission,
which traces dense gas of $\sim$ 10$^6$ cm$^{-3}$ (WWL95), and the fact that
none of the submillimetre peaks is coincident with any of the near-infrared
sources identified by Mendoza et al. (1990), suggests that the submillimetre
sources are very young objects, most likely Class 0 sources and/or prestellar
cores.
Similar anti-correlations between infrared and millimetre peaks are seen in
other star-forming regions (e.g., Casali, Eiroa, \& Duncan 1993;
Hurt \& Barsony 1996), and have been interpreted as reflecting different
phases of star formation.

% SCUBA maps for IRAS 12.
\begin{figure*}
% \includegraphics[width=168mm]{fig1.eps}
% \vspace{3.5cm}
 \caption{Contoured greyscale images (Jy/beam) of IRAS 12
  at {\bf a.} 850 $\mu$m: contours incremented in 0.24 Jy/beam intervals
  beginning with 0.24 Jy/beam;
  and {\bf b.} 450 $\mu$m: contours incremented in 1.2 Jy/beam intervals
  beginning with 1.2 Jy/beam.
% levels are at 30, 40, 50, 60, 70, 80, \& 90\% of the peak value; the 
  Submillimetre sources are marked with crosses; the
  {\it IRAS} source error ellipse is indicated; the positions of the
  near-infrared
  sources are marked with stars; the VLA source is marked with a triangle;
  and the H$_2$O maser is marked with a box.}
 \label{Submillimetre images of IRAS 12.}
\end{figure*}

\subsubsection{{\it IRAS} 06382+1017 (IRAS 25)}

Figure 2a shows
850 $\mu$m emission associated with IRAS 25. The position of the
{\it IRAS} source error ellipse, near-infrared source (star), Herbig--Haro knots
HH 124 A--F (xs),
and two VLA sources (triangles)
are indicated. Although the emission is extended, at least
two compact peaks can be identified (crosses).
These peaks are also apparent in the 450 $\mu$m emission shown in Figure 2b.
The brighter peak (25 S1) coincides approximately with VLA 2, which lies 
$\approx$15 arcsec to the east of VLA 1 
(approximately twice the 7 arcsec
resolution of our 450 $\mu$m SCUBA data). Unlike VLA 1, VLA 2 has
no point-source near-infrared counterpart. It lies in a concavity
seen at the north-east end of the infrared reflection nebulosity
reported by Pich\'e et al. (1995).
The slightly fainter SCUBA peak to the south-east of 25 S1 (25 S2)
is not associated with any
known near-infrared or radio continuum source.

\begin{figure*}
% SCUBA maps for IRAS 25.
% \includegraphics[width=168mm]{fig2.eps}
% \vspace{3.5cm}
 \caption{Contoured greyscale images (Jy/beam) of IRAS 25
  at {\bf a.} 850 $\mu$m: contours incremented in 0.20 Jy/beam intervals
  beginning with 0.30 Jy/beam;
  and {\bf b.} 450 $\mu$m: contours incremented in 0.7 Jy/beam intervals
  beginning with 1.1 Jy/beam.
  Submillimetre sources are marked with crosses; the {\it IRAS} source error
  ellipse is indicated, position of the near-infrared source is marked with a
  star, the VLA sources are marked with triangles, and HH 124 A--F are marked
  with xs.}
 \label{Submillimetre images of IRAS 25.}
\end{figure*}

\subsubsection{{\it IRAS} 06381+1039 (IRAS 27)}

Figure 3a shows the 850 $\mu$m
emission associated with IRAS 27. The position of
the {\it IRAS} source error ellipse is indicated.
We detected four compact submillimetre sources in the mapped region
(crosses). Three of these sources lie along a filament of extended
emission, similar to those seen in many other submillimetre maps of 
star-forming regions (e.g. Mitchell et al. 2001).
Figure 3b shows the 450 $\mu$m
emission associated with IRAS 27.
Although the image is very noisy, we were able to obtain fluxes for all
four compact sources.

\begin{figure*}
% SCUBA maps for IRAS 27.
% \includegraphics[width=168mm]{fig3.eps}
% \vspace{3.5cm}
 \caption{Contoured greyscale images (Jy/beam) of IRAS 27
  at {\bf a.} 850 $\mu$m: contours incremented in 0.32 Jy/beam intervals
  beginning with 0.32 Jy/beam;
  and {\bf b.} 450 $\mu$m: contours incremented in 1.2 Jy/beam intervals
  beginning with 1.2 Jy/beam.
  Submillimetre sources
  are marked with crosses and the {\it IRAS} source error
  ellipse is indicated.}
 \label{Submillimetre images of IRAS 27.}
\end{figure*}

\subsection{HIRES maps \& models}

In this section, we demonstrate that {\it IRAS} fluxes tend to probe the overall
properties of clumps -- protostellar groups -- rather than the properties
of individual protostellar cores. We show 
% In this section, we present FRESCO (Full-RESolution
% CO-add: equivalent to a single iteration of HIRES)
% maps for all four wavebands, and
% HIRES results achieved
% after 40 and 200 iterations for the 12 \& 25 $\mu$m and
% 60 \& 100 $\mu$m wavebands, respectively,
% for each region.
that HIRES point-source models based
on the source fluxes and positions given in the {\it IRAS} Point Source
Catalogue are completely inadequate to model the actual emission. Using 
% Note the separations of the submillimetre sources
% relative to the resolution  of the single-iteration {\it IRAS} images.
our SCUBA data to help guide placement and fluxes of input spikes, we
construct far more accurate models of the far-infrared
emission. Spike positions and fluxes are adjusted
until the models most closely match
the data in morphology and flux.
Figures 4--6 present HIRES maps for the three regions in the following
manner: the first row of each figure shows the FRESCO (Full-RESolution CO-add:
equivalent to a single iteration of HIRES) map for each waveband; the second
shows HIRES results achieved
after 40 iterations for the 12 \& 25 $\mu$m
wavebands, and after 200 iterations for the 60 \& 100 $\mu$m
wavebands; the third row shows point-source models of the HIRES-processed
emission using single spikes corresponding to the position and fluxes
given in the {\it IRAS} Point Source
Catalogue; and the fourth row shows the final models of the HIRES-processed
emission using multiple spikes.
No 12 $\mu$m maps are shown for IRAS 27, which was marginally detected at the
PSC position.
The most striking differences between the HIRES PSC and multiple spike models
are seen for IRAS 12, where objects contained within the {\it IRAS} source
are spread over a larger region than objects contained within the other
{\it IRAS} sources (figure 4).

Table 2 lists the final positions and magnitudes of HIRES model input spikes 
that were used for the relevant
wavebands, and the resolutions achieved at these positions. 
The associated errors are computed from 
the $\sim$20\% error inherent in HIRES (Levine \&
Surace 1993) in combination with the error associated with the point-source
models, which we estimate to be $<$8\%. This is the maximum amount
that the input fluxes can vary without causing significant changes in
the resulting models.
% For undetected sources, the upper limit presented corresponds to 
% $\sim 3\sigma$ detection.
We note that in all of the modelled regions
it was necessary to add additional spikes in at least one of the wavebands that
are not coincident with any known sources in order to closely simulate the
observed emission. We interpret this as being due to the presence of extended
dust emission in these regions, as is clearly seen in the SCUBA maps.
We emphasize that the HIRES-derived
source fluxes should therefore be interpreted
as upper limits to the actual source fluxes.
Additionally, many of the submillimetre sources were only weakly detected
at 12 \& 25 $\mu$m. HIRES processing is, in general, unreliable for unresolved
sources significantly fainter than 1 Jy (IPAC User's Guide, ed. 5).

The most accurate set of HIRES fluxes was derived for 12 S1 due to the
brightness of this object and its relative isolation from other nearby objects.
It was possible to model the HIRES-processed emission extremely closely
with a single spike located at the submillimetre source position.
We present the SED and derived properties
for this object in \S 3.5.
We note here that the Two Micron All Sky Survey (2MASS)
Point Source Catalogue (PSC) has been released since
completion of this work. Six of the submillimetre sources presented in this
paper have one or more faint near-infrared
counterparts. We will present SEDs and a more detailed analysis of these
sources in a subsequent paper.

\begin{table*}
\centering
\begin{minipage}{170mm}
 \caption{HIRES Model Spike Positions \& Fluxes}
 \begin{tabular}{lccrrrrrrrr}
  \hline
  Source 
  & $\alpha$(2000) & $\delta$(2000) &
  \multicolumn{4}{c}{Spike Height (Jy) at $\lambda$}  &
  \multicolumn{4}{c}{Effective Beam Size
  (Maj.$^{\prime\prime}\times$Min.$^{\prime\prime}$)}   \\
    &  &  &  
  \multicolumn{1}{c}{12 $\mu$m}  &  \multicolumn{1}{c}{25 $\mu$m}
  &  \multicolumn{1}{c}{60 $\mu$m}  &  \multicolumn{1}{c}{100 $\mu$m} &
  \multicolumn{1}{c}{12 $\mu$m}  &  \multicolumn{1}{c}{25 $\mu$m}
  &  \multicolumn{1}{c}{60 $\mu$m}  &  \multicolumn{1}{c}{100 $\mu$m} \\
  \hline
  RNO & 06$^h$41$^m$02.8$^s$   & 09$^{\circ}$36$^{\prime}$10$^{\prime\prime}$ 
  &  5.5$\pm$1.2  &  5.5$\pm$1.2  & 91$\pm$20  &  135$\pm$29
  & 47$\times$37  &  33$\times$29  &
  48$\times$41  &  51$\times$48  \\
   & 06$^h$41$^m$01.7$^s$   & 09$^{\circ}$36$^{\prime}$16$^{\prime\prime}$ 
  &   \dots  &  5.5$\pm$1.2  & \dots  &  \dots & \dots  &  33$\times$30  &
  \dots  &  \dots \\
  12 S1 & 06$^h$41$^m$05.8$^s$ & 09$^{\circ}$34$^{\prime}$09$^{\prime\prime}$ 
  &  $<$0.10  & 0.90$\pm$0.19  & 22$\pm$5 &  134$\pm$29
  & 73$\times$48 & 52$\times$42 &
  61$\times$47  &  86$\times$66  \\
  12 S2 & 06$^h$41$^m$06.2$^s$ & 09$^{\circ}$35$^{\prime}$57$^{\prime\prime}$ 
  &  0.30$\pm$0.06 & 1.1$\pm$0.2  & 27$\pm$6 &  74$\pm$16
  &  31$\times$28  &  31$\times$27  &
  32$\times$30  &  50$\times$45  \\
  12 S3 & 06$^h$41$^m$04.1$^s$ & 09$^{\circ}$35$^{\prime}$01$^{\prime\prime}$ 
  &  $<$0.10 & $<$0.10  &  6.9$\pm$1.5  &  39$\pm$8
  &  32$\times$21  &  35$\times$31  &
  57$\times$46  &  33$\times$33  \\
  12 S4 & 06$^h$41$^m$00.9$^s$ & 09$^{\circ}$35$^{\prime}$28$^{\prime\prime}$
  &  0.50$\pm$0.11 & 0.10$\pm$0.02   &  28$\pm$6  &  161$\pm$35
  &  33$\times$30  &  33$\times$29  &
  48$\times$40  &  49$\times$48  \\
  12 S5 & 06$^h$40$^m$49.3$^s$ & 09$^{\circ}$34$^{\prime}$36$^{\prime\prime}$ 
  &    $<$0.10 & 0.10$\pm$0.02  &  1.0$\pm$0.2  &  0.50$\pm$0.11
  &  70$\times$46  &  42$\times$28  &
  68$\times$31  &  17$\times$17  \\
  12 S6 & 06$^h$40$^m$57.8$^s$ & 09$^{\circ}$36$^{\prime}$25$^{\prime\prime}$ 
  &   0.70$\pm$0.15 & 0.60$\pm$0.13  & 16$\pm$3  &  10$\pm$2
  &  33$\times$30  &  31$\times$30  &
  45$\times$42  &  33$\times$32  \\
  12 S7 & 06$^h$41$^m$11.6$^s$ & 09$^{\circ}$35$^{\prime}$34$^{\prime\prime}$ 
  &   0.15$\pm$0.03  & 1.3$\pm$0.3   & 17$\pm$4  &  33$\pm$7
  &  70$\times$46  &  55$\times$44  &
  70$\times$49  &  67$\times$55  \\
  12 S8 & 06$^h$41$^m$09.2$^s$ & 09$^{\circ}$33$^{\prime}$01$^{\prime\prime}$ 
  &   $<$0.10  & 0.10$\pm$0.02   & 1.5$\pm$0.3  &  0.40$\pm$0.09
  &  59$\times$49  &  63$\times$51  &
  39$\times$29  &  33$\times$31  \\
  VLA & 06$^h$41$^m$04.5$^s$ & 09$^{\circ}$36$^{\prime}$20$^{\prime\prime}$ 
  & \dots & \dots   &  16$\pm$3  &  25$\pm$5  &  \dots  &  \dots  &
  32$\times$31 &  33$\times$32  \\
  IRS A & 06$^h$41$^m$06.5$^s$ & 09$^{\circ}$34$^{\prime}$44$^{\prime\prime}$
  & \dots & 1.5$\pm$0.3 &  \dots  &  \dots &  \dots  & 78$\times$63 &  \dots &
  \dots  \\
  IRS F & 06$^h$41$^m$09.0$^s$ & 09$^{\circ}$34$^{\prime}$08$^{\prime\prime}$
  & \dots & 0.70$\pm$0.15 &  \dots &  \dots &  \dots  & 50$\times$43 &  \dots &
  \dots  \\
    & 06$^h$41$^m$01.1$^s$   & 09$^{\circ}$36$^{\prime}$55$^{\prime\prime}$ 
  &   0.40$\pm$0.09 & 1.0$\pm$0.2   &  20$\pm$4  &  100$\pm$22
  &  34$\times$22 & 33$\times$22 & 32$\times$27 & 50$\times$46 \\
    & 06$^h$41$^m$01.6$^s$   & 09$^{\circ}$35$^{\prime}$49$^{\prime\prime}$ 
  & 2.0$\pm$0.4 & 3.2$\pm$0.7
  &  \dots  &  \dots & 49$\times$39 & 33$\times$27 & \dots & \dots \\
  25 NIR & 06$^h$41$^m$02.6$^s$ & 10$^{\circ}$15$^{\prime}$02$^{\prime\prime}$
  &   0.52$\pm$0.11  &  5.2$\pm$1.1  & 18$\pm$4  &  14$\pm$3
  &  51$\times$39  &  46$\times$29  &
  32$\times$31  &  51$\times$47  \\
  25 S1 & 06$^h$41$^m$03.5$^s$ & 10$^{\circ}$15$^{\prime}$10$^{\prime\prime}$ 
  &   0.17$\pm$0.04  & 1.9$\pm$0.4  & 9.5$\pm$2.0  &  22$\pm$5
  &  33$\times$27  &  32$\times$26  &
  32$\times$28  &  51$\times$47  \\
  25 S2 & 06$^h$41$^m$04.9$^s$ & 10$^{\circ}$14$^{\prime}$55$^{\prime\prime}$ 
  &  0.30$\pm$0.06  & 1.1$\pm$0.2  & 8.5$\pm$1.8  &  28$\pm$6
  &  32$\times$30  &  29$\times$28  &
  31$\times$30  &  50$\times$47  \\
   & 06$^h$41$^m$03.6$^s$  & 10$^{\circ}$14$^{\prime}$34$^{\prime\prime}$ 
  &   0.29$\pm$0.06   & \dots   & 8.0$\pm$1.7  &  \dots 
  & 50$\times$37  & \dots  & 32$\times$27
  & \dots  \\
  27 S1 & 06$^h$40$^m$58.5$^s$ & 10$^{\circ}$36$^{\prime}$54$^{\prime\prime}$
  &  0.25$\pm$0.05  & 2.5$\pm$0.5   &  35$\pm$8   &  73$\pm$16
  &  56$\times$42  &  55$\times$31  &
  32$\times$31  &  50$\times$48  \\
    & 06$^h$40$^m$59.0$^s$   & 10$^{\circ}$37$^{\prime}$04$^{\prime\prime}$ 
  &  \dots & \dots   &  18$\pm$4  &  \dots &  \dots &  \dots &
  32$\times$29  &  \dots \\
    & 06$^h$40$^m$58.0$^s$   & 10$^{\circ}$37$^{\prime}$04$^{\prime\prime}$ 
  &   \dots & \dots  & \dots  &  58$\pm$12  &  \dots &  \dots &
  \dots  &  50$\times$47  \\
  27 S2 & 06$^h$40$^m$59.1$^s$ & 10$^{\circ}$36$^{\prime}$09$^{\prime\prime}$
  &  $<$0.10  &  $<$0.10   &   7.0$\pm$1.5  &  13$\pm$3
  &  56$\times$45  &  33$\times$31  &
  44$\times$41  &  33$\times$32  \\
  27 S3 & 06$^h$41$^m$02.0$^s$ & 10$^{\circ}$35$^{\prime}$30$^{\prime\prime}$ 
  &  $<$0.10  &  0.10$\pm$0.02  &   1.2$\pm$0.3  &  $<$0.10 
  &  72$\times$43  &  74$\times$50  &
  59$\times$38  &  32$\times$27  \\
  27 S4 & 06$^h$40$^m$54.4$^s$ & 10$^{\circ}$33$^{\prime}$27$^{\prime\prime}$ 
  &  $<$0.10  &  $<$0.10   &  1.2$\pm$0.3  &  0.33$\pm$0.07
  &  60$\times$43  &  50$\times$41  &
  65$\times$41  &  51$\times$45  \\
   & 06$^h$41$^m$02.8$^s$   & 10$^{\circ}$35$^{\prime}$01$^{\prime\prime}$ 
  &  \dots & \dots & 1.2$\pm$0.3  &  \dots  & \dots  & \dots  &  52$\times$40
  & \dots  \\
  \hline
 \end{tabular}
\end{minipage}
\end{table*}
  
\subsubsection{{\it IRAS} 06382+0939 (IRAS 12)}

Figure 4 presents HIRES maps and models of the four {\it IRAS} wavebands
for IRAS 12. Very
little structure is present in all of the FRESCO maps (Figs. 4a--d); however,
the HIRES-processed maps (Figs. 4e--h) clearly indicate that not all of the
emission comes from {\it IRAS} 06382$+$0939: the 12 $\mu$m map shows a clear
extension toward 12 S6; the 25 \& 60 $\mu$m maps
show additional extensions toward
12 S7 and separate peaks at 12 S1; and the 100 $\mu$m map shows a pronounced
extension toward 12 S1.
Comparing these maps to models based
on Point Source Catalogue positions and fluxes (Figs. 4i--l)
illustrates the inadequacy of these models to
reproduce the actual emission.
For the models shown in Figs. 4m--p,
spikes were used at the submillimetre source positions;
the {\it IRAS} point source position; and various other positions
that either correspond to other known sources, or simply help simulate
extended emission.
% [GRACE comment: Cohen's integrated 47 $\mu$m flux for NGC 2264-S
% is overbright compared to our 60 $\mu$m flux and 25 $\mu$m upper limit,
% but there is a near-IR source mentioned in the Mendoza paper, which lies
% about 30 arcsec east of IRAS 12 S, and may be contributing to the
% mid-IR flux (note Cohen's 47 $\mu$m peak for NGC 2264-S is shifted east
% compared
% to the 95 $\mu$m peak.). They used integrated fluxes over a fairly large area.
% Our SCUBA data are significantly higher resolution
% than their KAO data, and the point-source HIRES modelling allows more accurate
% flux estimates.]

\begin{figure*}
% IRAS 12 FRESCO, HIRES, PSC models, and SCUBA models.
% \includegraphics[width=168mm]{fig4.eps}
% \vspace{16cm}
 \caption{
 IRAS 12 HIRES-processed emission after 1 iteration 
 at (a) 12 $\mu$m (contour levels are 5, 10, 15, \& 20 MJy/Ster),
 (b) 25 $\mu$m (contour levels are 10, 20, 30, \& 40 MJy/Ster),
 (c) 60 $\mu$m (contour levels are 50, 100, 150, 200, 250, \& 300 MJy/Ster),
 (d) 100 $\mu$m (contour levels are 100, 200, 300, \& 400 MJy/Ster);
 after 40 iterations at (e) 12 $\mu$m (contour levels are 20, 50, \& 100 
 MJy/Ster) and
 (f) 25 $\mu$m (contour levels are 20, 50, 100, \& 200 MJy/Ster);
 and after 200 iterations at (g) 60 $\mu$m (contour levels are 200, 600,
 1$\times$10$^3$,
 1.4$\times$10$^3$, 1.8$\times$10$^3$, \& 2.2$\times$10$^3$ MJy/Ster) and
 (h) 100 $\mu$m (contour levels are
 400, 800, 1.2$\times$10$^3$, 1.6$\times$10$^3$,
 2.0$\times$10$^3$, 2.4$\times$10$^3$, \& 2.8$\times$10$^3$ MJy/Ster).
 HIRES-processed models using PSC positions and fluxes after 40 iterations
 at (i) 12 $\mu$m (contour levels same as (e))
 and (j) 25 $\mu$m (contour levels same as (f));
 and after 200 iterations at
 (k) 60 $\mu$m (contour levels same as (g))
 and (l) 100 $\mu$m (contour levels same as (h)).
 Multi-spike HIRES-processed models after 40 iterations
 at (m) 12 $\mu$m (contour levels same as (e))
 and (n) 25 $\mu$m (contour levels same as (f)); and after 200 iterations at
 (o) 60 $\mu$m (contour levels same as (g))
 and (p) 100 $\mu$m (contour levels same as (h)).
 Source symbols are as follows: {\it IRAS} source position (ellipse);
 near-infrared sources (stars); submillimetre sources (large crosses);
 VLA source (triangle); and H$_2$O maser (box). 
 Small crosses mark the position of spikes not corresponding to any known
 sources that were used to more accurately simulate the emission.}
 \label{HIRES maps and models of IRAS 12.}
\end{figure*}

\subsubsection{{\it IRAS} 06382+1017 (IRAS 25)}

Figure 5 presents HIRES maps and models of the four {\it IRAS} wavebands
for IRAS 25.
Once again, very little structure is seen in the FRESCO maps (Figs. 5a--d),
except for a
pronounced shift to the northeast in the peak of the 100 $\mu$m emission.
The HIRES-processed {\it IRAS} emission (Figs. 5e--h) is well-peaked on
the {\it IRAS} source at 25 \& 60 $\mu$m, but the
100 $\mu$m emission clearly shows the peak to
be closer to the centre of 25 S1 \& 25 S2.
Models based on
Point Source Catalogue positions and fluxes (Figs. 5i--l)
fail to 
reproduce the actual emission, particularly at 100 $\mu$m.
Figs. 5m--p are HIRES models using spikes
at the submillimetre sources positions and
the near-infrared source position (coincident with the {\it IRAS} point
source).
An additional spike at no known source position was required to more accurately
model the 12 \& 60 $\mu$m emission.

\begin{figure*}
% IRAS 25 FRESCO, HIRES, PSC models, and SCUBA models.
% \includegraphics[width=168mm]{fig5.eps}
% \vspace{3.5cm}
 \caption{
 IRAS 25 HIRES-processed emission after 1 iteration 
 at (a) 12 $\mu$m (contour levels are 2, 2.5, 3, \& 3.5 MJy/Ster),
 (b) 25 $\mu$m (contour levels are 5, 10, \& 15 MJy/Ster),
 (c) 60 $\mu$m (contour levels are 10, 20, 30, 40, \& 50 MJy/Ster),
 (d) 100 $\mu$m (contour levels are 10, 15, \& 20 MJy/Ster);
 after 40 iterations at (e) 12 $\mu$m (contour levels are 5, 10, 15 \& 20
 MJy/Ster) and
 (f) 25 $\mu$m (contour levels are 20, 100, \& 200 MJy/Ster);
 and after 200 iterations at (g) 60 $\mu$m (contour levels are 100, 300, 500,
 \& 700 MJy/Ster) and
 (h) 100 $\mu$m (contour levels are
 100, 300, \& 500 MJy/Ster).
 HIRES-processed models using PSC positions and fluxes after 40 iterations
 at (i) 12 $\mu$m (contour levels same as (e))
 and (j) 25 $\mu$m (contour levels same as (f));
 and after 200 iterations at
 (k) 60 $\mu$m (contour levels same as (g))
 and (l) 100 $\mu$m (contour levels same as (h)).
 Multi-spike HIRES-processed models after 40 iterations
 at (m) 12 $\mu$m (contour levels same as (e))
 and (n) 25 $\mu$m (contour levels same as (f)); and after 200 iterations at
 (o) 60 $\mu$m (contour levels same as (g))
 and (p) 100 $\mu$m (contour levels same as (h)).
 Source symbols are as follows: {\it IRAS} source position (ellipse);
 near-infrared source (star); submillimetre sources (large crosses);
 VLA sources (triangles); and Herbig--Haro Objects (xs). The small cross in the
 12 \& 60 $\mu$m maps
 marks the position of a spike not corresponding to any known
 source that was used to more accurately simulate the emission.}
 \label{HIRES maps and models of IRAS 25.}
\end{figure*}

\subsubsection{{\it IRAS} 06381+1039 (IRAS 27)}

Figure 6 presents HIRES maps and models 
of the 25, 60, \& 100 $\mu$m
wavebands for IRAS 27. Since
27 S1 was marginally detected at 12 $\mu$m and the other sources
are non-detections, we do not include these maps.
There is little structure in the FRESCO maps (Figs. 6a--c). While the
HIRES-processed emission is peaked at the {\it IRAS} source position in
all wavebands (Fig. 6d--f),
the 60 $\mu$m map shows a clear extension along the arc formed
by 27 S1, 27 S2, \& 27 S3, and a separate, albeit weak,
peak appears close to the position
of 27 S4.
The Point Source Catalogue models (Figs. 6g--i) do not adequately reproduce this
emission. More accurate models (Figs. 6j--l) use spikes at the positions
of the submillimetre sources, as well as
multiple spikes in the immediate vicinity of 27 S1, and an additional spike
along the extended ridge of
60 $\mu$m emission.

\begin{figure*}
% IRAS 27 FRESCO, HIRES, PSC models, and SCUBA models.
% \includegraphics[width=148mm]{fig6.eps}
% \vspace{3.5cm}
 \caption{
IRAS 27 HIRES-processed emission after 1 iteration 
 (a) 25 $\mu$m (contour levels are 2, 3, 4, \& 5 MJy/Ster),
 (b) 60 $\mu$m (contour levels are 20, 30, 40, 50, \& 60 MJy/Ster),
 (c) 100 $\mu$m (contour levels are 30, 40, 50, 60, \& 70 MJy/Ster);
 after 40 iterations at 
 (d) 25 $\mu$m (contour levels are 10, 20, 30, 40, 50, 60, \& 70 MJy/Ster);
 and after 200 iterations at (e) 60 $\mu$m (contour levels are 20, 50, 100,
 400, 700, \& 1$\times$10$^3$ MJy/Ster) and
 (f) 100 $\mu$m (contour levels are
 100, 300, 500, \& 700 MJy/Ster).
 HIRES-processed models using PSC positions and fluxes after 40 iterations
 at (g) 25 $\mu$m (contour levels same as (d));
 and after 200 iterations at
 (h) 60 $\mu$m (contour levels same as (e))
 and (i) 100 $\mu$m (contour levels same as (f)).
 Multi-spike HIRES-processed models after 40 iterations
 at (j) 25 $\mu$m (contour levels same as (d)); and after 200 iterations at
 (k) 60 $\mu$m (contour levels same as (e))
 and (l) 100 $\mu$m (contour levels same as (f)).
 Source symbols are as follows: {\it IRAS} source position (ellipse)
 and submillimetre sources (large crosses).
 Small crosses mark the position of spikes not corresponding to any known
 sources that were used to more accurately simulate the emission.}
 \label{HIRES maps and models of IRAS 27.}
\end{figure*}

\subsection{CO Outflow Maps}

In this section, we present CO J=2$\rightarrow$1 maps of outflows
associated with the regions
containing IRAS 12, IRAS 25, \& IRAS 27.
A single outflow has been previously associated with each {\it IRAS} source
(NGC 2264 D, NGC 2264 O, \& NGC 2264 H), based on CO J=1$\rightarrow$0 and
CS J=2$\rightarrow$1 maps at half the resolution of the maps we present here
(MLS88; WW95).
Our maps indicate far more complex outflow morphology, and
suggest there are multiple outflows in each region.

\subsubsection{{\it IRAS} 06382+0939 (IRAS 12)}

Figure 7 shows blueshifted and redshifted high-velocity
CO J=2$\rightarrow$1 emission associated with outflow NGC 2264 D (contours)
superposed on the SCUBA 850 $\mu$m image (greyscale).
Submillimetre sources (crosses);
H$_2$O maser (square); VLA source (triangle); 
near infrared sources (stars);
and the {\it IRAS} error ellipse are indicated.
It is clear
that the {\it IRAS} point source, associated with the RNO binary,
is not associated with the high-velocity CO emission.
Unfortunately, we can not unambiguously
discern whether the submillimetre source,
12 S1, or the near-infrared source (IRS A) identified by Mendoza et al. (1990),
is the principal contributor to the outflow(s).
The major portions of the blue- and red-shifted lobes appear to lie at
different position angles. The blue lobe does not show a well-defined position
angle. The major component of the blueshifted emission extends almost due
north from the near-infrared source IRS F (Mendoza et al. 1990); to the
northeast of this component lies a finger of emission that points back
toward 12 S1, along a
P.A.$\sim$33 degrees. The major component of the redshifted emission
lies roughly along P.A.$\sim$80 degrees, although
a small finger of redshifted emission
extends just southwest of 12 S1 along a
position angle similar to the finger of blueshifted emission.
Clearly, more sensitive and higher-resolution
observations are needed to sort out possible contributions from the
various sources in this region, but there is no apparent CO outflow
associated with the {\it IRAS} point source.

\begin{figure}
% CO J=2-1 map of outflow NGC 2264 D.
% \includegraphics[width=84mm]{fig7.eps}
% \vspace{3.5cm}
 \caption{
Blueshifted (dashed contours) and redshifted (solid contours)
CO J=2$\rightarrow$1 emission associated with outflow NGC 2264 D
superposed on the SCUBA 850 $\mu$m image (greyscale: Jy/beam) of IRAS 12.
The blueshifted emission has been
integrated from -12.2 to 1.5 km s$^{-1}$ and is plotted in
10\% increments from 45\% to 95\% of the 31.2 K km s$^{-1}$ emission peak. 
The redshifted emission has been
integrated from 10.6 to 21.3 km s$^{-1}$ and is plotted in
10\% increments from 45\% to 95\% of the 25.6 K km s$^{-1}$ emission peak.
Submillimetre sources (crosses),
H$_2$O maser (square), VLA source (triangle), 
near infrared sources (stars) identified by
Mendoza et al. (1990), and the {\it IRAS} error ellipse are indicated.
Sources explicitly referred to in the text are labelled.
The beam size for the CO observations is indicated in the lower left corner.}
 \label{NGC 2264 D.}
\end{figure}

\subsubsection{{\it IRAS} 06382+1017 (IRAS 25)}

Figure 8 shows blueshifted and redshifted
CO J=2$\rightarrow$1 emission associated with outflow NGC 2264 O (contours)
superposed on the SCUBA 850 $\mu$m image (greyscale).
Submillimetre sources (crosses),
VLA sources (triangles), near-infrared source (star),
HH 124 A--F (xs),
and the {\it IRAS} source error
ellipse are indicated.
Pich\'e et al. (1995) suggested the possibility of two 
outflows in this region: one associated with HH 124, lying nearly
E--W, and another associated with a wider-angle
reflection nebula along a P.A.=45 degrees.
Our observations indicate three peaks in the high-velocity
CO emission: one peak in the redshifted emission and two peaks in the
blueshifted emission.
The western blueshifted lobe may be associated with VLA 1
(coincident with the {\it IRAS} source)
and the E--W Herbig--Haro flow, or it may be associated with 25 S1 and the
redshifted lobe to the north of 25 S1.
% The reflection nebula aligns fairly well with our CO red lobe and western
The southeastern blue lobe is probably not associated with the
Herbig--Haro flow, since the eastern
HH knot E is redshifted, while western knots A--D are strongly blueshifted
(Walsh et al. 1992). It may be associated with 25 S2.
In any case, the eastern blue lobe does not appear to be related to either of
the previously-reported outflows, which suggests there are
at least three outflows in the region.
Given the compactness of the outflows, it is not
possible to unambiguously associate outflows with sources at our resolution.

\begin{figure}
% CO J=2-1 map of outflow NGC 2264 O.
% \includegraphics[width=84mm]{fig8.eps}
% \vspace{3.5cm}
 \caption{
Blueshifted (dashed contours) and redshifted (solid contours)
CO J=2$\rightarrow$1 emission associated with outflow NGC 2264 O
superposed on the SCUBA 850 $\mu$m image (greyscale: Jy/beam) of IRAS 25.
The blueshifted emission has been
integrated from -0.2 to 4.9 km s$^{-1}$ and is plotted in
10\% increments from 60\% to 90\% of the 23.8 K km s$^{-1}$ emission peak. 
The redshifted emission has been
integrated from 11.1 to 19.2 km s$^{-1}$ and is plotted in
10\% increments from 25\% to 95\% of the 9.8 K km s$^{-1}$ emission peak.
Submillimetre sources (crosses), VLA sources (triangles), near-infrared source
(star) identified by Pich\'e et al. (1995), HH 124 A--F (xs),
and the {\it IRAS} source error
ellipse are indicated.
Sources explicitly referred to in the text are labelled.
The beam size for the CO observations is indicated in the lower left corner.}
 \label{NGC 2264 O.}
\end{figure}

\subsubsection{{\it IRAS} 06381+1039 (IRAS 27)}

Figure 9 shows blueshifted and redshifted
CO J=2$\rightarrow$1 emission associated with outflow NGC 2264 H (contours)
superposed on the SCUBA 850 $\mu$m image (greyscale). Submillimetre sources
(crosses) and the {\it IRAS} error ellipse are indicated.
The high velocity gas has a very complex morphology, but our map
helps to elucidate the ``squat'' appearance
of this outflow in
the MLS88 CO J=1$\rightarrow$0 map, where the ``major'' axis of the outflow
was reported to be shorter than the axis perpendicular to the outflow.
% Examination of the
% velocity structure in the line profiles suggests that the blueshifted emission
MLS88 identified this as a single outflow oriented in a north-south direction,
but our map indicates that the double-lobed redshifted emission 
lies east of 27 S1, along
P.A.$\sim$113 degrees,
and is probably
associated with this source
(which is also coincident with the {\it IRAS} source).
The small blue lobe directly
to the northwest of 27 S1 is probably part of this outflow, although the 
morphology, as well as the velocity structure, of the blue- and redshifted gas
are very different.

The origin of the blueshifted gas to the northeast of 27 S1 is
unclear, but may be elucidated in Figure 10, which shows blueshifted
emission integrated from 0.0 to 1.5 km s$^{-1}$ -- just outside of the line core
emission, and
redshifted emission integrated
from 8.5 to 10.0 km s$^{-1}$ -- just inside the line core, and hence
contaminated by ambient emission. Nevertheless,
the extended redshifted gas to the southwest and the blueshifted gas to the 
northeast are roughly bipolar about 27 S2 at a P.A.=35 degrees, suggesting
this source may power an outflow along this direction.
The ``finger'' of blueshifted emission extending northeast from 27 S2 lends
support to this interpretation.
The blueshifted gas directly northeast of 27 S3 may be associated 
with this source, but the emission is fairly weak and there is no 
strong evidence for redshifted gas on the opposite side of 27 S3.
In any case, the apparent reason for the ``squat'' appearance of NGC 2264 H
in the lower-resolution CO J=1$\rightarrow$0 maps (MLS88) is the presence
of at least two overlapping outflows in this region.
It is interesting to note that the outflows are approximately perpendicular
to the curved ridge of submillimetre emission that connects 27 S1, 27 S2,
and 27 S3.

\begin{figure}
% High-velocity CO J=2-1 map of outflow NGC 2264 H.
% \includegraphics[width=84mm]{fig9.eps}
% \vspace{3.5cm}
 \caption{
Blueshifted (dashed contours) and redshifted (solid contours)
CO J=2$\rightarrow$1 emission associated with outflow NGC 2264 H
superposed on the SCUBA 850 $\mu$m image (greyscale: Jy/beam) of IRAS 27.
Blueshifted emission from 0.0 to 3.0 km s$^{-1}$ is plotted as contour levels
in 10\% increments from 25\% to 95\% of the 12.7 K km s$^{-1}$ emission peak. 
Redshifted emission from 10.0 to 30.0 km s$^{-1}$ is plotted as contour
levels in 10\% increments from 25\% to 95\% of the 37.7 K km s$^{-1}$ emission
peak.
Submillimetre sources (crosses) and the {\it IRAS} error
ellipse are indicated.
Sources explicitly referred to in the text are labelled.
The beam size for the CO observations is indicated in the lower left corner.}
 \label{NGC 2264 H High-velocity.}
\end{figure}

\begin{figure}
% Low-velocity CO J=2-1 map of outflow NGC 2264 H.
% \includegraphics[width=84mm]{fig10.eps}
% \vspace{3.5cm}
 \caption{
Integrated emission from 0.0 to 1.5 km s$^{-1}$ is plotted (dashed contours) in
10\% increments from 20\% to 90\% of the 9.6 K km s$^{-1}$ emission peak.
Integrated emission from 8.5 to 10.0 km s$^{-1}$ is plotted (solid contours) in
10\% increments from 35\% to 95\% of the 28.6 K km s$^{-1}$ emission peak.
Submillimetre sources (crosses) and the {\it IRAS} error
ellipse are indicated.
Sources explicitly referred to in the text are labelled.
The beam size for the CO observations is indicated in the lower left corner.}
 \label{NGC 2264 H Low-velocity.}
\end{figure}

\subsection{IRAS 12 S1: An Intermediate-Mass Class 0 Object}

As discussed in \S 3.3, IRAS 12 S1 is the only source that could closely be
modelled with a single spike in all {\it IRAS} wavebands, with good spatial
separation from other sources. With the combined HIRES, SCUBA, and millimetre
fluxes, it is possible to plot a SED
with several reliable flux points on both
the Wien and Rayleigh-Jeans sides of the peak to help constrain physical
parameters of this object. Figure 11 displays this SED,
along with a single-temperature,
modified blackbody fit of the form:

\begin{equation}
S_{\nu}\ =\ B_{\nu}(T_d)(1\ -\ e^{-\tau_{\nu}})\, d\Omega
\end{equation}

\noindent
The best fit was derived using
a $\nu^{1.4}$ 
wavelength dependence of the dust optical depth.
The best-fit source diameter, d$\Omega$,
dust temperature, T$_d$, and 250 $\mu$m optical depth
are listed in Table 3. 
The source properties that can be derived from the model fit,
such as the source bolometric luminosity,
L$_{bol}$, and the circumstellar mass, M$_{env}$, are also
listed in Table 3, and were derived in a manner analogous to
that described in Barsony et al. (1998) for other Class 0 sources.
The tabulated bolometric luminosity was derived by
numerical integration under the fitted curve plotted in Fig. 11.

Class 0 objects have ratios of
$L_{submm}/L_{bol}\ \ge\ 5\ \times\ 10^{-3}$,
where $L_{submm}$ is the luminosity radiated 
longward of 350 $\mu$m (Andr\'e, Ward-Thompson, \& Barsony 1993).
The derived ratio of
$L_{submm}/L_{bol}$=0.03 for IRAS 12 S1 places this object well within the
Class 0 category.
Additionally, if IRAS 12 S1 is the main contributor to outflow NGC 2264
D, then its outflow momentum flux (1.3$\times 10^{-3}$
M$_{\odot}$ km s$^{-1}$ yr$^{-1}$ $\le$ F$_{CO}$ $\le$ 7.5$\times 10^{-3}$:
MLS88) is roughly an order of magnitude greater than expected for a Class I
object of comparable
bolometric luminosity, which is typical for Class 0 objects
(see, for e.g., fig. 5 in Bontemps et al. 1996).
The computed parameters suggest that IRAS 12 S1 comprises one or more
intermediate-mass protostars.

As stated in \S 3.3, the recently-released 2MASS PSC indicates that six of
the submillimetre sources discussed in this paper have near-infrared
counterparts. IRAS 12 S1 has
three faint near-infrared counterparts that
lie within a 7 arcsec radius of the SCUBA position.
Although lack of detection in the near-infrared has generally been used as
a criterion for Class 0 status, we note that 8 of the 42 
confirmed Class 0 objects listed in
Table 1 of Andr\'e, Ward-Thompson, \& Barsony (2000) can also be associated
with 2MASS point sources that lie within 7 arcsec of these objects:
W3OH-TW, L1448-N(A), L1641-VLA1, HH24MMS, IRAS 08076, L483-MM, G34.24$+$0.13MM,
\& S106-SMM. W3OH-TW \& G34.24$+$0.13MM are candidate massive Class 0 objects,
and L1641-VLA1, IRAS 08076, \& L483-MM are listed as borderline Class 0
objects (Andr\'e et al. 2000). L1448-N(A) has recently been suggested
to be a borderline Class 0 object by O'Linger et al. (2003), who argue that lack
of detection shortward of 10 $\mu$m should be discarded as a criterion for
Class 0 status, since this characteristic reflects current technology rather
than intrinsic source properties.

\begin{figure}
% SED for IRAS 12 S1.
% \includegraphics[width=84mm]{fig11.eps}
% \vspace{3.5cm}
 \caption{
SED of IRAS 12 S1: Plotted fluxes are presented in Tables 1 \& 2. Parameters
of the plotted fit are presented in Table 3.
}
 \label{IRAS 12 S1 SED.}
\end{figure}

\begin{table}
 \caption{IRAS 12 S1 Source Properties}
 \begin{tabular}{@{}lcc}
  \hline
  Parameter & Symbol \& Units & Value \\
  \hline
  IRAS 12 S1 $\alpha$(2000) &  & 06$^h$41$^m$05.8$^s$  \\
  IRAS 12 S1 $\delta$(2000) &  & 09$^{\circ}$34$^{\prime}$09$^{\prime\prime}$ \\
  Adopted Distance & D(pc) & 800 \\
  Fit temperature & T$_d$ (K) & 23 \\
  Fit optical depth & $\tau_{250 \mu m}$  &  0.04  \\
  Fit source diameter & d$\Omega$ (arcsec) & 28  \\
  Bolometric luminosity & L$_{bol}$ (L$_{\odot}$) & 107.5  \\
       & L$_{submm}$/L$_{bol}$  &  0.03  \\
  Circumstellar envelope mass & M$_{env}$ (M$_{\odot}$) & 17.6 \\
  \hline
 \end{tabular}
 
 \medskip
The adopted distance, originally estimated by Walker (1956), is the same
distance adopted for the surveys of Margulis, Lada, \& Snell (1988) and
Margulis, Lada, \& Young (1989), which are discussed in the text.
\end{table}

\section
{Discussion}

It is important to ascertain which sources within clumps 
drive molecular outflows and/or jets in order to investigate similarities
and differences between low- and high-mass YSOs
that are associated with
these phenomena, and to accurately correlate sources and outflow parameters.
Our new observations and modelling indicate that
previously categorized ``Class I objects''
in the Mon OB1 dark cloud are, in fact, associated with
protoclusters containing multiple
sources at different evolutionary stages.
In two cases, the object identified as the
``{\it IRAS} point source'' is undetected at submillimetre wavelengths, but
lies near multiple submillimetre sources.
In both cases, the submillimetre sources contribute significantly to the
{\it IRAS} fluxes at 60 \& 100 $\mu$m, but the {\it IRAS} source
dominates at shorter wavelengths.

Correlations between outflow force or momentum deposition rate, F$_{CO}$, and
source luminosity, L$_{bol}$, have
been noted over many decades of luminosity
since the earliest surveys for outflows (e.g., Bally \& Lada 1983;
Panagia 1991; Cabrit \& Bertout 1992).
Many of these correlations were established by
making extensive use of the {\it IRAS} database to help
establish source luminosities.
Furthermore,
the F$_{CO}$/L$_{bol}$ ratio has been used to estimate the fraction
of the accretion flow that is ejected in the wind, in order to help distinguish
between different models of the wind ejection mechanism (e.g., Richer et al.
2000).
Our results suggest that
many of the noted source--outflow correlations may reflect the sum total of
source and outflow properties from a protocluster or protogroup rather than the
properties of individual sources within the cluster.
Hence, the relationship between outflow momentum and source bolometric
luminosity at high source luminosities is particularly called into question.

To investigate the effect of our results on the previously-calculated {\it IRAS}
luminosities for the sources in this study,
we note that
MLY89 calculated {\it IRAS} luminosities for the Mon OB1 sources IRAS 12,
IRAS 25, \& IRAS 27 to be 330 L$_{\odot}$, 110 L$_{\odot}$, \& 87 L$_{\odot}$,
respectively.
Applying MLY89's equation (1) to the {\it IRAS} upper limits we derived for
the ``point sources'', we find the luminosities for IRAS 12 (RNO),
IRAS 25 (NIR), and IRAS 27 (27 S1) to be 150 L$_{\odot}$, 28 L$_{\odot}$, \&
74 L$_{\odot}$, respectively. For the first two cases the recalculated
luminosities are significantly lower than the original calculations. 
Calculated in this manner, 
the {\it IRAS} luminosity of the source that we
identify as a likely contributor to NGC 2264 D, 12 S1, is only
$\sim$48 L$_{\odot}$; however, the bulk of the luminosity from this source
is emitted at longer wavelengths and our calculated bolometric luminosity is
107.5 L$_{\odot}$.
% Only
% in the case of IRAS 27 is the recalculated luminosity not significantly
% different. IRAS 27 is also the only region where the {\it IRAS} peak
% and strongest submillimetre peak coincide, with the most luminous source
% being the principal contributor to the very young outflow, NGC 2264 H.

Our results indicate that extreme care must be taken when using outflow
momentum deposition rates and source luminosities in order to infer
the relationship between accretion rates and outflow.
Outflow momentum deposition rates cannot be unequivocally correlated
with source luminosities in many of the clustered regions for which
these luminosities were derived from {\it IRAS} fluxes.
% Additionally, our results suggest that many ``Class I'' sources that are 
% located in confused regions may comprise multiple sources in the
% prestellar, Class 0, 
% and/or Class I phase of evolution.
The similar resolving ability of the Stratospheric Observatory
for Infrared Astronomy (SOFIA) at infrared wavelengths
to the JCMT's SCUBA
at submillimetre wavelengths, combined with higher-resolution
outflow observations, should enable a critical assessment of the properties
of outflows and driving sources in many
clustered regions.

\section
{Summary}

\begin{enumerate}
\item
We have used HIRES-processing and modelling of {\it IRAS} data, along with SCUBA
imaging, to identify individual cores embedded in massive CS
clumps in the Mon OB1 dark cloud, and estimate their contributions to the 
{\it IRAS} fluxes.

\item
Each CS clump had been previously associated with a single
{\it IRAS} point source. The associated YSO was, in each case,
linked with a single molecular outflow and had been assigned a 
Class I type SED.

\item
In two of the three clumps studied, none of the bright submillimetre sources
is coincident with the identified
{\it IRAS} ``point source''. 

\item
In all three CS clumps, an associated {\it IRAS} point source
dominates the 12 \& 25 $\mu$m emission; however, new objects identified
through their submillimetre continuum emission and distinct from the 12 \&
25 $\mu$m {\it IRAS} point sources, are major contributors to the observed,
extended 60 \& 100 $\mu$m {\it IRAS} emission. These results suggest that the
previously classified ``Class I'' objects actually consist of multiple
sources at different evolutionary stages.

\item
We were able to closely model the object IRAS 12 S1 using 
a single spike in all {\it IRAS} wavebands, with good spatial
separation from other sources. Using the combined HIRES, SCUBA, and millimetre
fluxes to plot an SED
with several reliable flux points on both
the Wien and Rayleigh-Jeans sides of the peak, we have calculated the
dust temperature, bolometric luminosity, and circumstellar mass of this
source. The SED and physical parameters of this source, together with its
likely status as the principal driving source of outflow NGC 2264 D,
suggest that IRAS 12 S1 is a Class 0 Object harboring one or more
intermediate-mass protostars.

\item
While it is not possible to derive individual SEDs for all of the embedded
objects due to the presence of extended emission and source confusion,
we can at least place good
upper limits on the individual flux contributions and thus 
compare the new {\it IRAS} ``point source'' luminosities to luminosities
that were previously derived using {\it IRAS} Point Source Catalogue fluxes.
In the two cases where the {\it IRAS} point source is not coincident with
a submillimetre source, such a comparison indicates the {\it IRAS} source
to be of significantly lower bolometric luminosity than previously estimated.

\item
Our new CO J=2$\rightarrow$1 outflow maps
of the three regions we studied indicate complex outflow morphology suggestive
of multiple driving sources.
We find that the submillimetre source,
12 S1,  contributes to driving the NGC 2264 D outflow; however, the previously
assumed driving source for this flow, the source at the {\it IRAS} PSC
position (RNO), cannot be associated with any high-velocity CO emission.
There are hints of three separate outflows in the
vicinity of IRAS 25:
a giant Herbig--Haro flow, probably powered by the near-infrared source which
lies at the {\it IRAS} PSC position, 
and two very compact CO outflows that may be associated with the
submillimetre sources 25 S1 \& 25 S2.
% contribution from the near-infrared source.
In the IRAS 27 region, the strongest submillimetre source (27 S1), 
which is also coincident with the
{\it IRAS} PSC position, apparently drives the redshifted gas associated with
the outflow NGC 2264 H. The multi-lobed appearance of the blueshifted gas,
and the morphology of the low-velocity redshifted gas, suggest
the presence of two other outflows in this region: a bipolar 
outflow associated 27 S2, and blueshifted gas that may be associated with 27 S3.

\item
Our results have implications for the method of using outflow
momentum deposition rates and source bolometric luminosities in order to infer
the fraction of the accretion flow that is ejected into the wind of
protostars, and thus for using this method to distinguish between
different proposed launching mechanisms. It is clear that in many cases
outflow momentum deposition rates cannot be unequivocally correlated
with source luminosities in many of the clustered regions for which
luminosities were derived using {\it IRAS} fluxes.

\item
Six of the submillimetre objects identified in this paper can be associated
with faint objects identified from the recently-released Two Micron All
Sky Survey Point Source Catalogue. A detailed analysis of the physical 
properties and evolutionary status of these sources will be presented in a
subsequent paper.
\end{enumerate}

\section*{Acknowledgments}

A portion of this work was performed while G.W.-C.
held a President's Fellowship from the University of California.
M.B. gratefully acknowledges support through NSF grants AST 00-96087
(CAREER) and AST-0206146, which made her contributions to this work possible.
G.W.-C. wishes to thank Jose Francisco Salgado for his assistance with
Adobe Illustrator, which was used to produce several of the figures.
We also wish to thank the referee, Gary Fuller, for his helpful suggestions to
improve this paper.

This research has made use of the NASA/IPAC Infrared Science Archive, which is
operated by the Jet Propulsion Laboratory, California Institute of Technology,
under contract with the National Aeronautics and Space Administration.
This publication makes use of data products from the Two Micron All Sky
Survey, which is a joint project of the University of Massachusetts and the
Infrared Processing and Analysis Center/California Institute of Technology,
funded by the National Aeronautics and Space Administration and the National
Science Foundation.
% MB's NSF CAREER Grant,
% AST95-01788.

\end{document}